\documentclass[letterpaper]{article} 
\usepackage{aaai25}  
\usepackage{times}  
\usepackage{helvet}  
\usepackage{courier}  
\usepackage[hyphens]{url}  
\usepackage{graphicx} 
\usepackage{natbib}  
\usepackage{caption} 
\usepackage{algorithm}
\usepackage{amsmath}
\usepackage{algpseudocode}
\usepackage{booktabs}
\usepackage{fontawesome}
\usepackage{tikz}

\newcommand{\checkbox}{%
    \tikz\draw[thick](0,0) rectangle (0.2,0.2) -- (0.05,0.1) -- (0.1,0.05) -- (0.15,0.15);
}

\newcommand{\crossedbox}{%
    \tikz\draw[thick](0,0) rectangle (0.2,0.2) 
    -- (0.2,0.2) (0,0.2) -- (0.2,0);
}

\usepackage{newfloat}
\usepackage{listings}
\DeclareCaptionStyle{ruled}{labelfont=normalfont,labelsep=colon,strut=off} 
\floatstyle{ruled}
\newfloat{listing}{tb}{lst}{}
\floatname{listing}{Listing}
\title{ NeuroLit Navigator: A Neurosymbolic Approach to Scholarly Article Searches for Systematic Reviews}

\author {
     Vedant Khandelwal\textsuperscript{\rm 1},
     Kaushik Roy\textsuperscript{\rm 1},
     Valerie Lookingbill\textsuperscript{\rm 2},
     Ritvik Garimella\textsuperscript{\rm 1},
     Harshul Surana\textsuperscript{\rm 1},
     Heather Heckman\textsuperscript{\rm 2},
     Amit Sheth\textsuperscript{\rm 1}
 }
 \affiliations {
     \textsuperscript{\rm 1}Artificial Intelligence Institute, University of South Carolina\\
     \textsuperscript{\rm 2}Library Sciences, University of South Carolina\\
     kaushik@email.com, vedant@email.com, amit@email.com, ritvik@email.com, 
     harshul@email.com, valerie@email.com, heather@email.com
}

\usepackage{bibentry}

\begin{document}

\maketitle

\begin{abstract}
The introduction of Large Language Models (LLMs) has significantly impacted various fields, including education, for example, by enabling the creation of personalized learning materials. However, their use in Systematic Reviews (SRs) reveals limitations such as restricted access to specialized vocabularies, lack of domain-specific reasoning, and a tendency to generate inaccurate information. Existing SR tools often rely on traditional NLP methods and fail to address these issues adequately. To overcome these challenges, we developed the ``NeuroLit Navigator,'' a system that combines domain-specific LLMs with structured knowledge sources like Medical Subject Headings (MeSH) and the Unified Medical Language System (UMLS). This integration enhances query formulation, expands search vocabularies, and deepens search scopes, enabling more precise searches. Deployed in multiple universities and tested by over a dozen librarians, the NeuroLit Navigator has reduced the time required for initial literature searches by 90\%. Despite this efficiency, the initial set of articles retrieved can vary in relevance and quality. Nonetheless, the system has greatly improved the reproducibility of search results, demonstrating its potential to support librarians in the SR process. 

\faGithub~Code is available at \cite{anonymousCoder}
\end{abstract}

\section{Introduction}
Systematic reviews (SRs) are pivotal in academic research, underpinning scholarly endeavors by consolidating existing knowledge and identifying gaps in the literature. These reviews are crucial for maintaining scientific rigor, informing policy decisions, and guiding future research directions. However, conducting SRs is inherently laborious, often involving screening thousands of scholarly articles to determine their relevance to a research question. This task demands extensive research time and effort and poses a significant challenge in maintaining consistency and objectivity \cite{bullers2018takes}.
Large Language Models (LLMs) have emerged as powerful tools in various natural language processing applications, notably transforming educational technologies. These models, exemplified by systems like OpenAI's GPT series, excel in tasks ranging from automated content creation to personalized learning environments \cite{domenichini2024llms}. However, their deployment in conducting SRs reveals critical limitations. Despite their sophistication, LLMs often struggle with domain-specific reasoning and are restricted by the availability of database-specific controlled vocabulary \cite{gross2015still}. This is compounded by their tendency to produce ``hallucinated'' information—fabricating data or misrepresenting facts—which undermines their reliability \cite{rawte2023troubling}. A challenge highlighted in recent studies is the models' poor performance in accurately retrieving the correct citations of papers, a task crucial for maintaining the integrity of academic research \cite{tilwani2024reasons}. Such shortcomings underscore the need for enhanced methods that can effectively navigate the complexities of academic literature.

The key challenge in automating SRs is formulating effective search queries and retrieving relevant articles. Existing approaches often rely on keyword-matching, which frequently includes irrelevant articles while excluding pertinent ones. Additionally, the iterative refinement of search strategies based on initial findings adds complexity to automating the process \cite{bramer2018systematic}. Recent advances in AI, particularly Retrieval-Augmented Generation (RAG) models, have improved document retrieval in SRs. These models combine large-scale information retrieval with deep learning to predict relevance \cite{lewis2020retrieval}. However, RAG systems offer limited query refinement and struggle to integrate the evolving research context, often requiring several iterations to achieve satisfactory results \cite{yang2024rag}.


In response to these challenges, our work introduces the ``NeuroLit Navigator,'' an innovative system designed to optimize the first iteration of article retrieval in SRs. It takes user input queries along with sentinel articles, key papers, or studies highly relevant to the research question, serving as a starting point or benchmark for the search. By integrating multiple knowledge graphs (KGs) and a novel Neurosymbolic AI framework, our system enhances the query formulation process, enabling more sophisticated analysis and interpretation of search terms in relation to the vast body of academic literature. NeuroLit Navigator employs a combination of symbolic reasoning and neural learning techniques to understand and expand queries in a more informed and contextually relevant manner. It also incorporates controlled vocabulary, which involves standardized terms and phrases that help maintain consistency and precision in retrieving relevant literature. Using controlled vocabulary is considered best practice according to major systematic review guidelines, as it minimizes variability in search results and ensures comprehensive coverage of the topic \cite{chandler2019cochrane,JBI2024}. This approach not only emphasizes early-stage effectiveness, specifically during the initial iteration but also sets the groundwork for subsequent, more detailed iterations by researchers, who can refine and expand the search criteria based on the initial set of articles retrieved. This methodology acknowledges the inherent complexities of SRs and avoids overstating the system's capabilities, focusing instead on its role in enhancing the preliminary stages of article collection.



\section{Background}

\subsection{Systematic Reviews in Academic Research} 

SRs synthesize research findings systematically and comprehensively, which is crucial for upholding scientific integrity. SRs are pivotal in higher education as they ensure students and educators can access thoroughly vetted and relevant information. Assuming they are well-conducted and based on methodologically rigorous studies, SRs mitigate bias by methodically collecting and analyzing literature, in turn, offering one of the highest forms of evidence \cite{andersen2024comparative} to inform evidence-based decision-making, curriculum development, and pedagogical approaches across scientific disciplines. However, research has repeatedly shown that SR searches contain errors that impact precision and recall \cite{rethlefsen2015librarian,koffel2015use}, which can impact quality of a review’s conclusions.

As publication rates for SRs have steadily increased with an average yearly growth rate of 26\% \cite{andersen2024comparative}, it is critical to ensure the quality of SR searches to support well-informed decisions and teaching practices. Librarians, with their expertise in systematic searching, play a crucial role in helping students and educators develop search strategies that are thorough, reproducible, and aligned with best practices. However, developing comprehensive search strategies is time and labor-intensive, with query development averaging 8-15 hours \cite{bullers2018takes}. As the demand for and reliance on SRs continues to rise, so does the need for more efficient and accurate search processes. The introduction of NeuroLit Navigator marks a significant advancement in educational technology, specifically designed to assist academic librarians. Streamlining the search process enables librarians to support students and educators better in quickly integrating the latest scholarly findings into pedagogical strategies and curriculum development. This enhancement is essential for keeping educational programs relevant and effective in an era of rapid knowledge evolution and interdisciplinary demands, thus solidifying the role of librarians as key facilitators in the educational ecosystem.

\subsection{LLMs in Systematic Reviews}
LLMs have significantly advanced natural language processing but exhibit limitations when applied to SRs. LLMs often lack the precision needed for SRs due to restricted vocabularies and are limited by their training on public datasets \cite{gross2015still}. This results in suboptimal search and retrieval outcomes. Furthermore, LLMs may produce hallucinated content, complicating their use in SRs where accuracy is crucial \cite{rawte2023troubling}. The NeuroLit Navigator system addresses these issues by integrating domain-specific LLMs with structured knowledge sources like MeSH and UMLS, enhancing the relevance of search results. For more details, see Supplementary \ref{appendix:llms_in_srs}.


\subsection{Retrieval-Augmented Generation (RAG)}

Retrieval-augmented generation (RAG) combines retrieval-based methods with generative models to enhance the accuracy and relevance of generated responses \cite{lewis2020retrieval}. It utilizes existing information from large databases to inform its responses, ensuring they are grounded in real-world data. The RAG system operates through two main stages: (i) \textbf{Retrieval Step:} This step retrieves relevant data from external sources using various methods, ensuring the information is closely related to the input query. (ii) \textbf{Generation Step:} Using the retrieved data, the system then generates a response that is not only based on the query but also informed by the retrieved information. This method typically uses advanced models like Transformers but can be adapted to other models as needed. This combination allows for more accurate and contextually relevant responses by incorporating external knowledge into the generation process.

\subsubsection{Limitations in Systematic Reviews}
Despite its effectiveness in various applications, RAG has several limitations in the context of SRs: (i) \textbf{Information Hallucination:} RAG can sometimes create content that isn't supported by the retrieved data, especially when the data retrieved is only partially relevant or irrelevant. (ii) \textbf{Handling Evolving Datasets:} RAG can struggle to keep up with rapidly updating and growing datasets, which may hinder its ability to fetch the most recent and relevant documents. (iii) \textbf{First-Iteration Limitations:} The first iteration in SRs is critical. RAG may not effectively handle complex search criteria initially, necessitating enhancements for better precision. Our approach improves this by starting with a more sophisticated initial query formulation integrating structured knowledge from specialized areas.

These challenges require ongoing improvements to ensure both the accuracy of retrieval and the relevance of the generated content, especially in fields where precision is crucial.

\subsection{Knowledge Graphs and Neurosymbolic AI}
Neurosymbolic AI combines neural networks with symbolic reasoning to enhance learning and reasoning capabilities in systematic literature reviews \cite{sheth2023neurosymbolic}. While neural networks excel at recognizing patterns in large datasets, they often lack interpretability and require extensive data. On the other hand, Symbolic AI provides clear, rule-based reasoning but struggles with complex, unstructured data. Our system, Neurolit Navigator, leverages Neurosymbolic AI to improve the accuracy and relevance of SRs. By integrating structured KGs from biomedical ontologies and KGs, such as Medical Subject Headings (MeSH) \cite{mesh} and the Unified Medical Language System (UMLS) \cite{lindberg1990unified}, the system enhances interpretability, query formulation, and refinement. 

This integration bridges the gap between data-driven learning and rule-based reasoning, enabling neural networks to utilize structured knowledge for more precise query generation. For instance, in a query about “antimicrobial agents and surgical site infections,” MeSH expands the search vocabulary to include related concepts such as “anti-infective agents,” “therapeutic uses,” and “wound infections.” UMLS connects these broader terms to more specific ones, such as “antibacterial agents,” ensuring a comprehensive search that captures relevant studies beyond simple keyword searches. This integration of KGs like MeSH and UMLS enables our system to interpret and expand queries effectively, improving the breadth and accuracy of SRs. For additional examples and details on KGs, see Supplementary \ref{appendix:knowledge_graphs}.

\section{Existing works in systematic reviews}

Incorporating machine learning and artificial intelligence technologies has enhanced the SR process. This section reviews several prominent systems and identifies key areas for improvement that NeuroLit Navigator addresses.

Several existing systems for systematic literature reviews face common challenges related to handling domain-specific vocabularies, scalability, and the need for continuous human input. \textbf{Sysrev}, while leveraging FAIR principles to enhance collaboration and data reusability, lacks advanced natural language processing (NLP) capabilities and struggles with domain-specific vocabularies essential for specialized fields like biomedical research \cite{bozada2021sysrev}. Similarly, \textbf{FAST2} reduces initial human effort but still depends on user input for adapting to evolving research topics, limiting its flexibility \cite{yu2019fast2}. These systems are constrained by their inability to adjust to complex terminologies dynamically without human assistance. Other systems face challenges in balancing precision and recall. The \textbf{Cochrane Review} classifier aimed at identifying randomized controlled trials (RCTs) achieved high recall but suffered from low precision, resulting in the retrieval of irrelevant articles requiring manual screening \cite{thomas2021machine}. This issue also occurs in \textbf{Dextr}, which automates data extraction for public health literature but still requires human verification to ensure accuracy \cite{walker2022evaluation}. Both systems rely heavily on manual intervention to address their limitations in retrieval accuracy.

Moreover, systems like \textbf{Colandr}, \textbf{EPPI-Reviewer}, and \textbf{SWIFT-Active Screener} streamline SRs using machine learning but depend on iterative refinement, requiring significant human input \cite{cheng2021keep, eppi_reviewer_web_beta, howard2020swift}. While these systems improve document screening speed, their reliance on user interaction and model retraining makes it difficult to handle evolving datasets effectively without oversight. Other LLMs-based tools enhance scientific research by enabling users to query in plain language. Scite.ai categorizes citations to provide relevant references \cite{scite}, while Perplexity.ai combines indexing and LLM-based searches to pull pertinent web pages \cite{perplex}. Consensus Copilot searches over 200 million research papers using OpenAI's LLMs for comprehensive summaries \cite{consensus}. However, despite their innovative capabilities these tools struggle with a lack of domain-specific knowledge and controlled vocabulary. This often results in the retrieval of incorrect or irrelevant articles, a critical issue where the accuracy and precision are vital for dependable scientific outcomes.

SRs have increasingly incorporated machine learning and AI technologies to streamline the review process. Among these, \textbf{GEAR-Up} \cite{roy2024gear} represents a significant advancement by employing a modular approach that integrates Generative AI and external KGs to assist librarians. GEAR-Up processes queries through natural language processing, expanding them using external knowledge sources and generating related queries via language models like ChatGPT. It then narrows the search results using a FAISS-powered re-ranking algorithm, focusing on keywords extracted from user queries. However, it primarily relies on keyword-based retrieval and does not consider the broader context of sentinel articles, which can limit the depth and context of search results. \noindent While GEAR-Up significantly aids the query generation process and retrieves relevant articles, its reliance on keyword expansions, without using controlled vocabulary and a FAISS-powered retriever focuses on speed but may not always ensure the relevancy and breadth of the retrieved articles, focusing strictly on the keywords generated in earlier stages. Moreover, like NeuroLit Navigator, GEAR-Up focuses on optimizing the first iteration of the SR process, aiming to set a new baseline for efficiency and comprehensiveness in academic searches. More details are in Supplementary \ref{appendix:compare_gear}.

A key limitation of existing systems is their reliance on human input for setup and refinement, such as manual validation, training, or query adjustments. This limits scalability in handling complex domain-specific vocabularies. Unlike systems like GEAR-Up, our approach improves SRs through a zero-shot method that eliminates manual input for query formulation. By integrating domain-specific LLMs with structured KGs like MeSH and UMLS, NeuroLit Navigator captures complex biomedical terms and generates precise, context-aware queries. This fully automated, scalable solution enhances query development and retrieval, surpassing the limitations of systems like Sysrev and FAST2.

\begin{figure*}[t]
\centering
\includegraphics[width=0.75\textwidth]{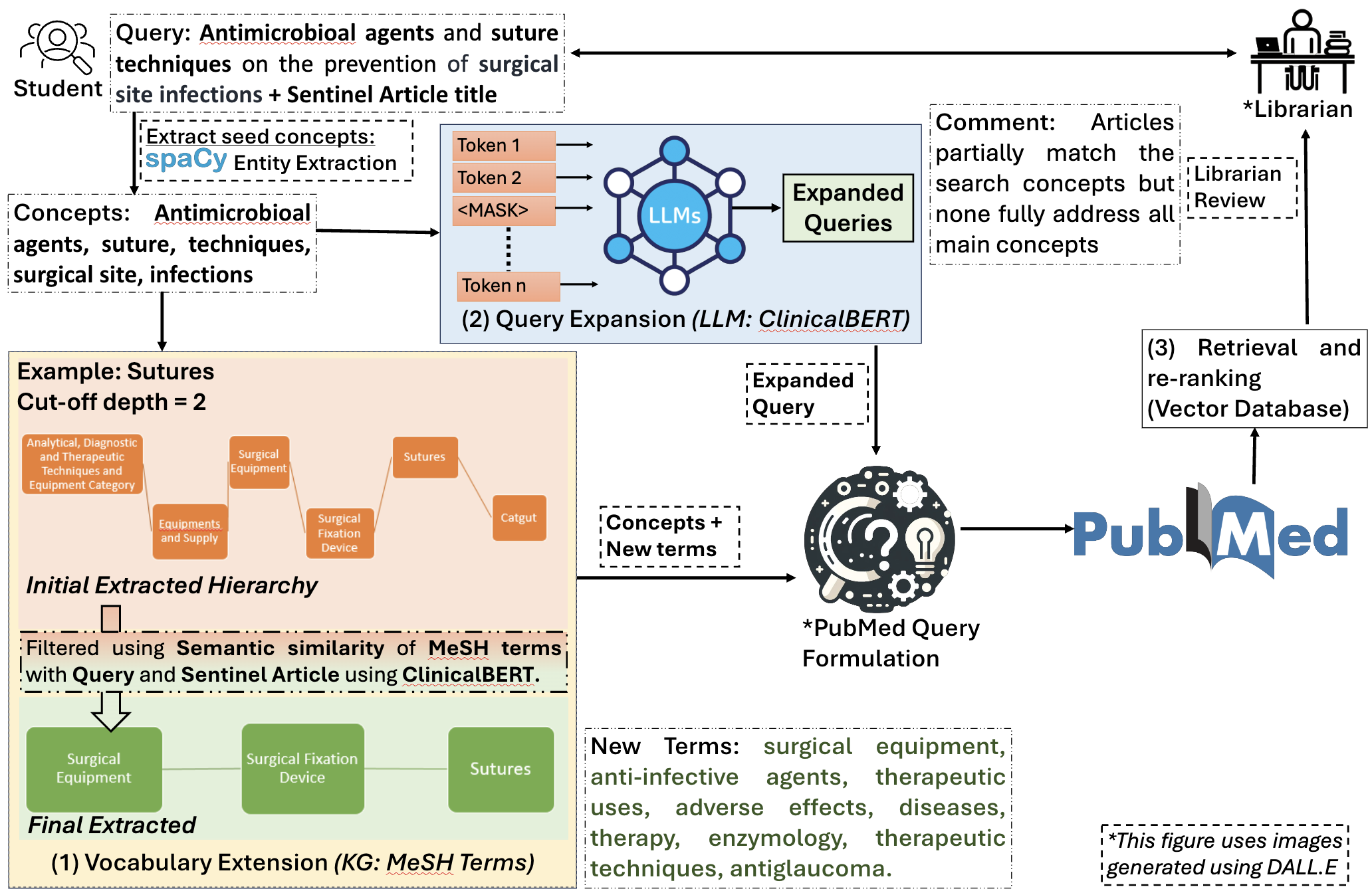}
\caption{The technical architecture demonstrates the three-step process in detail: (1) User input is processed through domain-specific Named Entity Recognition (NER) to extract key terms and phrases from the query and sentinel articles. (2) In parallel, Vocabulary Extension leverages KGs to expand relevant terms within two hops of the extracted concepts. At the same time, Query Expansion uses domain-specific LLMs (such as ClinicalBERT) to replace key terms with related substitutes. (3) The final step involves article retrieval and re-ranking based on semantic similarity to the user query, with the top 5 articles provided for librarian feedback.}
\label{fig:tech_arc}
\end{figure*}

\section{Methods}

The system enhances SRs by leveraging domain-specific knowledge and LLMs. The methodology (as shown in Figure \ref{fig:tech_arc}) follows a three-step process: user input, vocabulary extension and query expansion, and article retrieval and re-ranking. Each step is critical in refining the search query and the retrieval of relevant articles for systematic review.

\subsubsection{Step 1: User Input and Named Entity Recognition}

The process begins with the user providing two essential inputs: the research query and a sentinel article or a set of articles. The sentinel article represents the most relevant article(s) to the user’s research query. This is the gold standard against which all retrieved articles will be compared. Once the input is received, the system uses domain-specific Named Entity Recognition (NER) models to extract key terms and phrases from the query and the sentinel articles. These terms serve as the basis for the subsequent steps in the system, guiding vocabulary extension and query expansion.

\subsubsection{Step 2: Vocabulary Extension and Query Expansion}

After extracting key terms and phrases, the system proceeds with two parallel processes to refine and expand the query:

\paragraph{Vocabulary Extension:} Using domain-specific KGs (such as MeSH for biomedical terms), the system identifies related terms up to two hops away from the extracted key terms. This allows the query to capture broader but contextually relevant concepts that may not have been included in the original user query. For example, if the key term is "sutures," the knowledge graph might suggest related terms such as "surgical equipment" and "catgut." Semantic filtering is performed to identify related terms based on the similarity of all the two hops away terms in KGs with user input query and sentinel articles.

\paragraph{Query Expansion:} In parallel, the system uses domain-specific LLMs (such as ClinicalBERT) to mask key terms in the query and sentinel articles, generating relevant substitutes. These substitutes help expand the query to include synonymous or related terms that the user might not have explicitly included. For example, the system might replace "antimicrobial agents" with related terms such as "anti-infective agents" or "therapeutic techniques." These two processes—vocabulary extension and query expansion—generate an expanded set of terms that enrich the original query. After these steps, we employ a comprehensive process to build SQL query to query the databases. This query-building process is a dynamic and iterative procedure designed to optimize the retrieval of relevant articles. The system begins with highly specific queries and gradually broadens them when the initial search does not yield sufficient results. Controlled vocabulary terms, such as [Mesh] or [tiab], are applied to key terms within the query, ensuring consistency and enhancing precision by aligning the search with standardized terminology recognized in systematic reviews. This approach ensures that the system retrieves both precise and comprehensive results. Algorithm \ref{alg:querybuilding} describes the query refinement process.

\begin{algorithm}[!htb]
\caption{Iterative Query Refinement Process}
\begin{algorithmic}[1]
\State \textbf{Input:} User query $Q$, entities $E = \{e_1, e_2, \dots, e_n\}$, KG terms $M_i$ for each $e_i$, masks $K_i$, threshold $N_{\text{min}}$
\State \textbf{Output:} Final query $Q_{\text{final}}$, retrieved articles
\State Construct query: 
\[
Q_{\text{specific}} = \bigwedge_{i=1}^{n} \left( e_i \lor \bigvee_{m \in M_i} m \lor \bigvee_{k \in K_i} k \right)
\]
\State Retrieve $articles \gets \text{get\_data}(Q_{\text{specific}})$
\While{$\text{num}(articles) < N_{\text{min}}$ and $|E| > 0$}
    \State Remove least relevant entity: $E \gets E \setminus \{e_j\}$
    \State Rebuild query: 
    \[
    Q_{\text{general}} = \bigwedge_{i=1, i \neq j}^{n} \left( e_i \lor \bigvee_{m \in M_i} m \lor \bigvee_{k \in K_i} k \right)
    \]
    \State Retrieve $articles \gets \text{get\_data}(Q_{\text{general}})$
\EndWhile
\State \Return $Q_{\text{general}}, articles$
\end{algorithmic}
\label{alg:querybuilding}
\end{algorithm}

The process begins by constructing the most specific query \( Q_{\text{specific}} \) using entities extracted from the user query, their associated KG terms, and mask expansions. This ensures that the system starts with a highly precise search. If fewer than the desired number of articles \( N_{\text{min}} \) are retrieved, the system calculates relevance scores for each entity and removes the least relevant one. A more generalized query \( Q_{\text{general}} \) is then constructed, broadening the search scope. The system continues this iterative process until enough relevant articles are retrieved.

\subsubsection{Step 3: Retrieval and Re-Ranking}

Once the query has been fully expanded and built, it is used to query the database (such as PubMed) for article retrieval. The system retrieves a set of articles based on the expanded query. It proceeds to re-rank them based on their semantic similarity to the user’s input query and sentinel articles. This re-ranking process prioritizes the most relevant articles, optimizing retrieval. The system uses a vector database to compute the semantic similarity of each retrieved article to the expanded query. The top k articles are selected and presented to the user (librarian) for further evaluation and feedback. This Neurosymbolic AI-driven process ensures that the most contextually relevant articles are retrieved, allowing for an efficient and accurate SR process.

\section{Experimental Setup}

\subsection{Domain-Specific Resources}

In this study, we utilize a suite of domain-specific models to ensure that the NeuroLit Navigator system is finely tuned for the biomedical domain. Our approach involves evaluating several advanced embedding models, each pre-trained on domain-specific data crucial for handling the highly specialized vocabulary and concepts inherent in biomedical fields.

The models we assessed include Distill-Bert \cite{reimers-2019-sentence-bert}, MPNet \cite{biompnet}, PubMedQA \cite{deka2022improved}, SPECTER \cite{cohan2020specter}, and MedCPT \cite{jin2023medcpt}. These models were chosen for their potential to enhance the system's ability to process biomedical literature effectively. Our selection process was guided by the need to find an embedding model that best fits the specific requirements of the SR process in biomedical research. Among the models tested, we have opted to integrate the following into our system due to their particular strengths in dealing with biomedical data:

\begin{itemize}
    \item \textbf{ClinicalBERT:} A transformer-based model specifically pre-trained on clinical notes and biomedical literature, ClinicalBERT is utilized for enhancing the breadth of our search queries by suggesting semantically related terms based on masked input \cite{wang2023optimized}.
    
    \item \textbf{SciSpacy (en\_core\_sci\_lg):} This large biomedical-specific named entity recognition (NER) model extracts relevant concepts from scientific texts, facilitating precise inputs for the vocabulary extension and query expansion processes \cite{neumann-etal-2019-scispacy}.
    
    \item \textbf{MPNet Model:} After thoroughly testing several domain-specific models, we proceeded with the MPNet model. This model excels at computing semantic embeddings of queries and retrieved articles, particularly those from PubMed abstracts and citations. MPNet’s robust performance in re-ranking articles based on their relevance to the original query and sentinel articles is instrumental in achieving the high precision required by our system.

    \item \textbf{API:} For this work, we query the PubMed database and use PubMed Entrez API for querying the database and retrieving candidate articles.
\end{itemize}

Further, we have used k=5 to retrieve the top 5 articles given a user input query and sentinel article, which are further evaluated by domain experts.

\subsection{Deployment}

The NeuroLit Navigator system is being tested to be integrated into the library infrastructures of multiple universities. Librarians from these universities have tested the system in real-time, handling diverse academic queries from various disciplines. This dynamic deployment enables the system to be tested in a highly realistic environment where user interactions reflect actual academic search requirements. Each librarian inputs research queries and sentinel articles as part of the deployment. These inputs are processed through the Neurosymbolic pipeline, which includes NER, vocabulary extension, and query expansion, followed by article retrieval and re-ranking. The feedback loop is crucial in refining the system by allowing librarians to provide structured feedback on each search query's effectiveness.

\subsection{Feedback and Evaluation Mechanism}

The feedback mechanism is designed to capture the librarian’s evaluation of the relevance and completeness of the results provided by the system. After retrieving and re-ranking articles, the system displays the top-5 results, which are compared against the sentinel article(s). Librarians fill out a customized feedback form that allows them to:

\begin{itemize}
    \item \textbf{Identify Missing Concepts:} Librarians can indicate whether certain relevant terms or concepts were missing from the retrieved articles.
    \item \textbf{Assess Query Interpretation:} Librarians provide feedback on whether the system correctly interpreted the research query and captured the core ideas.
    \item \textbf{Rate Relevance:} Librarians score the relevance of each retrieved article in the top-5 results, focusing on whether the articles match the scope and depth of the input query.
\end{itemize}

This structured feedback is logged and used to compute the \textbf{Relevance \% metric}, which is the average ratio of relevant articles to the total number of articles retrieved across various input queries.

\subsection{Comparative Analysis}

To benchmark the NeuroLit Navigator system, we conducted a comparative analysis against several established retrieval tools and LLM-based systems. This comparison specifically focuses on the capabilities of existing systems to handle the initial iteration of the SR search process, a critical phase for setting the foundation of a systematic review. 

We directly compare NeuroLit Navigator with GEAR-Up, the current baseline for SRs focused on the first iteration. GEAR-Up, known for its integration of Generative AI with external knowledge, serves as a vital benchmark to gauge our system's advancements in handling complex biomedical queries more effectively. Additionally, we evaluate our system against other LLM-only retrieval approaches that primarily rely on generative language models for search queries without the use of any external knowledge sources. Each system is evaluated by domain experts, who assess the relevance and comprehensiveness of the articles retrieved. This evaluation allows us to demonstrate NeuroLit Navigator's superior capabilities in enhancing the precision and scope of SRs through the integration of advanced domain-specific knowledge and cutting-edge language model technologies.



    
    

\subsection{Ethical Considerations}

This research adheres to stringent ethical standards, ensuring user data privacy and confidentiality. All librarian feedback is anonymized, with no personally identifiable information collected. We comply with all relevant privacy laws and have received IRB approval, verifying our adherence to ethical guidelines. This structured approach enables a thorough evaluation of the NeuroLit Navigator system, highlighting its effectiveness and areas for improvement.

\section{Deployed System Working}

The deployed system is designed to enhance the efficiency of SR's through a user-friendly interface. Below, we describe the system's operational workflow, illustrated with screenshots from the actual interface. A sample example of input, output, and feedback is in Supplementary \ref{appendix:example}.

\subsection{System Interface and Initial Query Input}


\begin{figure}[ht]
\hspace*{-0.12\linewidth}
\rotatebox{270}{\includegraphics[width=0.9\linewidth]
{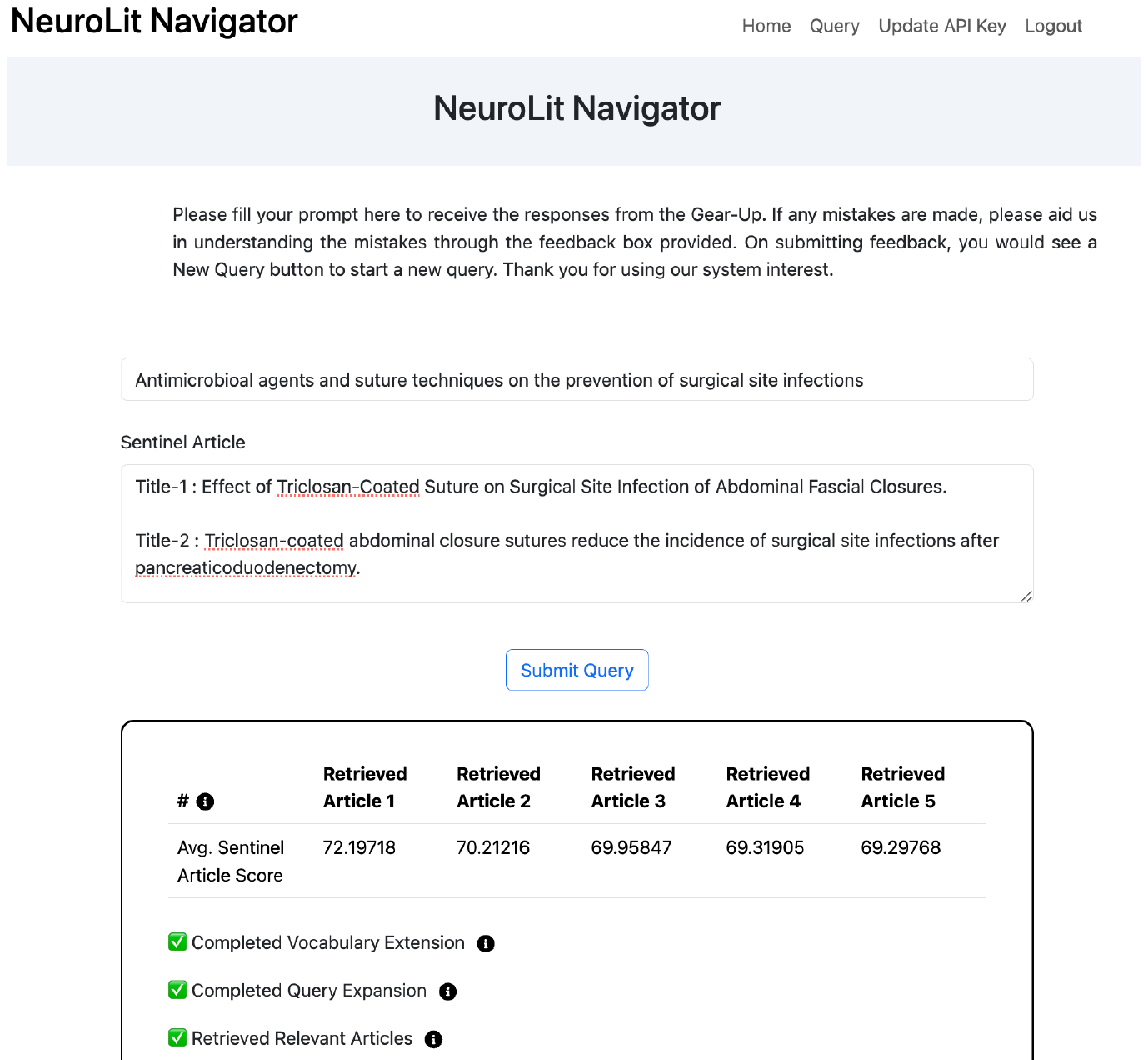}}
\caption{Updated initial input interface where users enter their research queries. This screen serves as the gateway for users to define the scope of their research by entering queries related to their systematic review, enhanced by the option to include a sentinel article for better context alignment.}
\label{fig:input_deploy_updated}
\end{figure}

The revised initial input interface (Figure \ref{fig:input_deploy_updated}) presents a streamlined, user-focused entry point for SRs. Users begin by entering specific search terms or complex queries related to their research topics in the provided text box. Additionally, they can specify a sentinel article, which aids the system in aligning the retrieved literature with the query context, thereby improving the precision of the search results.

Following the submission of the initial query, the system performs several background operations:
\begin{itemize}
    \item \textbf{Vocabulary Extension:} Automatically expands the search terms to include synonyms and related terms, ensuring comprehensive topic coverage.
    \item \textbf{Query Expansion:} Enhances the original query by incorporating additional relevant terms derived from the sentinel article and the expanded vocabulary, which helps in fetching more pertinent articles.
    \item \textbf{Article Retrieval:} Executes the refined query against large-scale academic databases to fetch relevant articles, with their relevance scores displayed alongside.
\end{itemize}

After performing this background process, the scores of the retrieved articles are similar to those of sentinel articles, as shown in Figure \ref{fig:input_deploy_updated}. This interface facilitates an intuitive and efficient start to the SR process. It incorporates advanced features that leverage the system’s AI capabilities to enhance the relevance and accuracy of search results.

\subsection{Retrieved Article}

\begin{figure}[ht]
\centering
\includegraphics[width=0.9\linewidth]{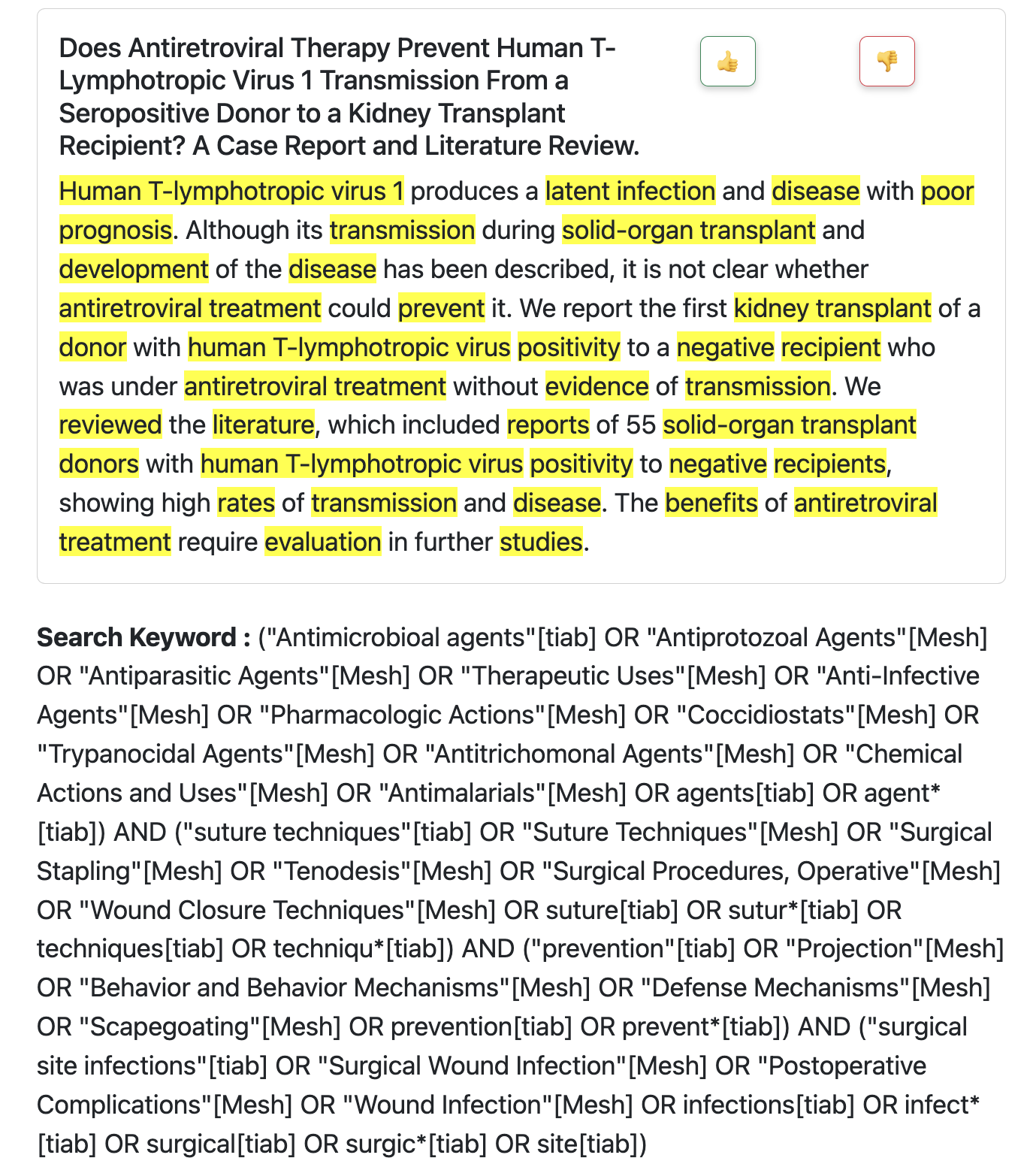}
\caption{Display of a retrieved article with key sections highlighted, and serach query. This feature assists users in quickly identifying the relevance of the article to their query, enhancing the review process efficiency.}
\label{fig:article_deploy}
\end{figure}

Once the query is processed, the system displays the retrieved articles. As shown in Figure \ref{fig:article_deploy}, articles are presented with key excerpts highlighted to aid users in swiftly assessing each document's relevance. This feature is crucial for rapid review and interpretability and ensures users spend less time on non-relevant literature. Further, it also shows the search query that was used to retrieve the article, allowing reproducibility of the system output.

\subsection{Feedback}
\begin{figure}[ht]
\centering
\includegraphics[width=0.9\linewidth]{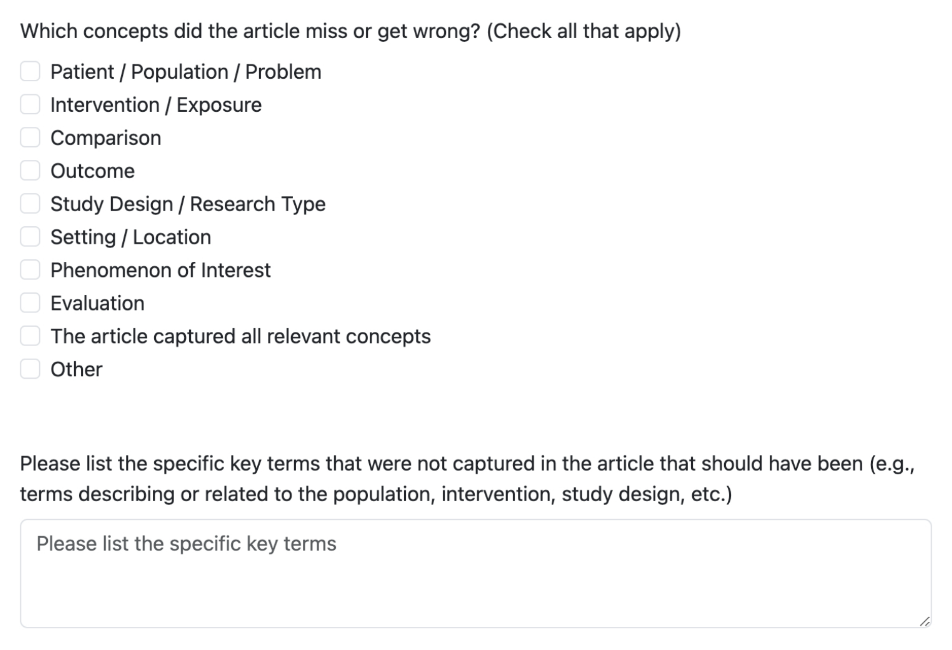}
\caption{Feedback form for each article. Users can indicate which concepts were missed or misrepresented, thus refining the system's future search accuracy and relevance.}
\label{fig:feedback_deploy}
\end{figure}

A feedback mechanism is integrated to further enhance the system’s accuracy and adaptiveness, as depicted in Figure \ref{fig:feedback_deploy}. After reviewing each article, users can provide specific feedback about missed or misrepresented concepts. This feedback is crucial for refining the system's search algorithms and ensuring continuous improvement in the retrieval and analysis of relevant literature.

Combining these features—query input, relevant article display with highlights, and feedback provision—ensures a robust system that supports the complex demands of SRs, making the review process more manageable and precise.

\section{Results}

The comparative evaluation of embedding models for the NeuroLit Navigator system highlighted significant performance differences. To identify the best-performing model, we computed the relevance percentage of retrieved articles with the help of domain experts. MPNet achieved the highest relevance percentage at 36\%, followed by BertDistill at 28\%. SPECTER and PubmedBert also achieve a relevance score of 24\% and 20\%, respectively. Finally, MedCPT becomes the worst performing with a relevance percentage of 4\%. More details are in Supplementary \ref{appendix:domainmodels}.

Transitioning from model performance to system evaluation, Table \ref{tab:system_comparison} contrasts NeuroLit Navigator with other systems using criteria crucial for systematic reviews (SRs).

\begin{table}[!htb]
\centering
\setlength{\heavyrulewidth}{1.5pt} 
\begin{tabular}{p{2cm} p{1.8cm} p{.6cm} p{.6cm} p{.6cm} p{.6cm}}
\toprule
\textbf{System} & \textbf{Relevance\%} & \textbf{R} & \textbf{I}  & \textbf{CV} \\ 
\midrule
Scite           & 33\%                         & No                       & No                                   & No                           \\ 
Consensus       & 38\%                         & No                       & No                                 & No                           \\ 
Perplexity      & 33\%                         & No                       & No                                  & No                           \\ 
GEAR-Up         & 26.6\%                       & No                       & Yes                                   & No                           \\ 
\textbf{NeuroLit Navigator} & \textbf{36\%} & \textbf{Yes} & \textbf{Yes}  & \textbf{Yes} \\ 
\bottomrule
\end{tabular}
\caption{Comparison of various systems based on relevance percentage, reproducibility (R), interpretability (I), and use of controlled vocabulary (CV). NeuroLit Navigator outperforms other systems in relevance and is the only system providing both reproducibility and interpretability without inherent bias, along with employing controlled vocabulary, which enhances search precision and consistency in SRs.}
\label{tab:system_comparison}
\end{table}

As shown in Table \ref{tab:system_comparison}, while NeuroLit Navigator does not achieve the highest relevance percentage, it excels in combining high relevance with reproducibility and interpretability, features crucial for systematic reviews (SRs). The slightly lower relevance percentage of 36\%, compared to 38\% of Consensus, is offset by the benefits that NeuroLit Navigator offers. This includes the integration of a controlled vocabulary, which standardizes terminology and minimizes variability in search results. This feature is particularly valuable in SRs, where the precision and consistency of term usage are essential for producing reliable and replicable findings \cite{chandler2019cochrane,JBI2024}. Furthermore, the absence of inclusion of reproducibility confirm the superiority of NeuroLit Navigator in providing dependable results. Output from LLM-based tools is discussed in Supplementary \ref{appendix:llmoutputs}.

\section{Conclusion and Future Work}

The ``NeuroLit Navigator'' has made significant progress in automating systematic literature searches by combining domain-specific LLMs with structured knowledge sources. This integration enhances query formulation, improves precision and recall, and reduces the time required for initial searches by 90\% in real-world academic applications. However, challenges remain, such as difficulties in identifying emerging research areas not covered by structured resources and the need for human input at later review stages. The system’s effectiveness depends on the continuous adaptation of LLMs to evolving research trends.

Future improvements will focus on multi-iteration refinement and user feedback integration, allowing the system to automatically adjust search parameters based on input and results. This will enhance responsiveness and reduce dependence on static knowledge sources, particularly in rapidly changing fields, further streamlining the review process.

\section*{Acknowledgements}
This research extends previous work cited in \cite{roy2024gear}, partially supported by NSF Award 2335967 "EAGER: Knowledge-guided neurosymbolic AI" for safe virtual health assistants. Author opinions, not sponsor opinions, are expressed \cite{sheth2021knowledge,sheth2022process,sheth2023neurosymbolic,sheth2024neurosymbolic,sheth2024neurosymbolicb,sheth2024civilizing,sheth2025c}.

\appendix

\section{LLMs in Systematic Reviews}
\label{appendix:llms_in_srs}

Large Language Models (LLMs) have significantly influenced advancements in natural language processing but exhibit notable shortcomings when applied to Systematic Reviews (SRs). These models, trained on extensive and diverse datasets, often struggle with the precision needed for SRs due to their limited access to domain-specific vocabularies \cite{gross2015still}. This deficiency results in a mismatch between the specialized terminology inherent in scientific literature and the generalist vocabulary understood by LLMs, leading to suboptimal search and retrieval results. Furthermore, LLMs are constrained by their training on public datasets. They may not have direct access to comprehensive, proprietary academic databases, further complicating their ability to fetch relevant and up-to-date scientific articles.

Another critical issue is their tendency to produce hallucinated content—fabricated information unsupported by the source material \cite{rawte2023troubling}. This characteristic is particularly problematic in SRs, where the accuracy and reliability of information are paramount. \cite{tilwani2024reasons} underscores a significant flaw in LLMs concerning their ability to accurately generate citations, often resulting in citations that do not align with the discussed scientific concepts, thereby misleading the review process. LLMs sometimes fail to grasp the complex, nuanced discussions typical in academic writings, leading to oversimplified or contextually incorrect interpretations.

These limitations highlight the need for a tailored approach to integrate LLMs' extensive generative capabilities with precise, structured methodologies required in systematic reviews. Our NeuroLit Navigator system addresses these issues by integrating domain-specific Large Language Models with structured knowledge graphs such as MeSH and UMLS. This combination enriches the query development process and ensures the search outputs are precise, contextually accurate, and aligned with domain-specific requirements. NeuroLit Navigator leverages zero-shot capabilities to adapt dynamically to evolving research topics without extensive retraining, thus maintaining high standards of accuracy and relevance in Systematic Reviews.

\section{Detailed Discussion on Knowledge Graphs in Systematic Reviews}
\label{appendix:knowledge_graphs}

Knowledge Graphs (KGs) are expert-curated, structured compilations of interpretable symbolic concepts and relationships between them, representing the symbolic component of our neurosymbolic analysis. These graphs embody ground truth conceptual knowledge and facilitate structured querying and systematic analysis, which are crucial for enhancing AI's capability to structure data to mirror human understanding and logical deduction \cite{sheth2023neurosymbolic}.

Our system utilizes well-established biomedical KGs such as Medical Subject Headings (MeSH) \cite{mesh} and the Unified Medical Language System (UMLS)\cite{bodenreider2004unified}. MeSH structures biomedical concepts hierarchically and is used to index articles in PubMed and other databases, facilitating both broad and narrow searches. UMLS integrates over 200 biomedical vocabularies, linking related concepts across different terminologies, which enriches the connections between search terms.

An example of this is a query for ``sutures used in liver transplant procedures.'' By leveraging the hierarchical structure of MeSH, the system identifies related broader terms such as ``surgical equipment.'' UMLS then connects these terms with specific materials like ``catgut sutures,'' making the query more comprehensive. This advanced search process ensures that results from databases like PubMed cover a wider spectrum of related studies, offering more robust and relevant articles than a traditional keyword-based search.

These knowledge graphs significantly improve the comprehensiveness of systematic reviews, allowing researchers to access a broader array of relevant articles, thereby enhancing the search process's recall and precision.

\section{Comparison with GEAR-Up}\label{appendix:compare_gear}
GEAR-Up employs a modular approach to assist librarians with systematic reviews, enhancing query development by integrating Generative AI and external knowledge graphs. Initially, it processes queries through natural language processing to expand them using external knowledge sources, followed by generating related queries via language models like ChatGPT. Finally, it narrows the search results using a FAISS-powered retriever, focusing on keywords extracted from the user queries without considering any sentinel articles \cite{roy2024gear}.

While GEAR-Up significantly aids the query generation process and retrieves relevant articles, it primarily relies on keyword expansions. It does not consider the broader context or depth provided by sentinel articles. Moreover, its use of FAISS for article retrieval focuses on speed but may not always ensure the relevancy and breadth of the retrieved articles, as it strictly adheres to the keywords generated in earlier stages.

In contrast, NeuroLit Navigator significantly enriches query formulation by integrating domain-specific knowledge graphs, LLMs, and advanced embedding models. It incorporates multiple keywords and contextual cues from user queries and sentinel articles, enhancing the scope and relevance of searches. NeuroLit Navigator's re-ranking and retrieval system uses domain-specific embedding models for a more semantically aware article selection, ensuring greater relevancy and reproducibility. This system focuses on optimizing the first iteration of the systematic review process, setting a new baseline for efficiency and comprehensiveness in academic searches.

These enhancements make NeuroLit Navigator fundamentally stronger and more adaptable to the complex needs of systematic literature reviews, significantly advancing beyond the capabilities demonstrated by GEAR-Up.

\section{Example of System Output} \label{appendix:example}
The NeuroLit Navigator systematically processes input queries to retrieve relevant scientific abstracts. Below is an illustrative example demonstrating how the system handles a query and the corresponding outputs.

\subsection{Input and Output}

\subsubsection{Input Question}
\noindent\textbf{Research Query (RQ):} Gender affirming surgeries for female-to-male transgender individuals.

\noindent\textbf{Sentinel Articles:} Title: Female-to-Male Gender-Affirming Chest Reconstruction Surgery; Journal: Aesthetic surgery journal.

\noindent\textbf{Retrieved Abstracts Overview:} Each abstract is retrieved based on a comprehensive search key, specifically tailored to the input query, focusing on multiple related terms to ensure a broad yet accurate coverage.

\noindent\textbf{Search Key:}
\begin{lstlisting}[breaklines=true, basicstyle=\small\ttfamily, numbers=none]
("Gender"[tiab] OR Gender[tiab] OR gender*[tiab]) AND 
("surgeries"[tiab] OR surgeries[tiab] OR surgeri*[tiab]) AND 
("female-to-male transgender"[tiab] OR female[tiab] OR femal*[tiab] OR 
transgender[tiab] OR transgend*[tiab] OR Gender[tiab] OR gender*[tiab]) AND 
("individuals"[tiab] OR individuals[tiab] OR individu*[tiab])
\end{lstlisting}

\noindent\textbf{Retrieved Abstracts:}

\begin{enumerate}
    \item \textbf{Title:} Gender-affirming surgeries for transgender and gender diverse individuals and associated health outcomes.\\
    \textbf{Similarity Score:} 81.43\\
    \textbf{Abstract:} Gender-affirming care is treatment that recognizes and affirms the gender identity of transgender and gender-diverse (TGD) individuals. This review describes the most common procedures considered to be gender-affirming surgeries (GAS), detailing important relevant considerations for each procedure. Patient outcomes and the legal landscape for GAS are also discussed.

    \item \textbf{Title:} Urological focus on gender affirmation surgery.\\
    \textbf{Similarity Score:} 78.23\\
    \textbf{Abstract:} Gender affirmation surgery plays an important role in the treatment of gender dysphoria. This article provides an overview of genital gender affirmation surgeries, focusing on the available options and appropriate referral criteria for general practitioners and surgeons.

    \item \textbf{Title:} Isolated and combined breast augmentation in transgender patients: Multi-institutional insights into early outcomes and risk factors.\\
    \textbf{Similarity Score:} 73.12 \\
    \textbf{Abstract:} This study analyzes the outcomes of breast augmentation surgery (BAS) for female transgender individuals, evaluating safety profiles and identifying risk factors for adverse events.

    \item \textbf{Title:} Anesthesia for gender-affirming surgery: a practical review.\\
    \textbf{Similarity Score:} 72.66\\
    \textbf{Abstract:} This review discusses the relevant anesthetic considerations for the most common gender-affirming surgeries (GAS), which often involve highly specialized surgical techniques.

    \item \textbf{Title:} Sexual Function in Post-surgical Transgender and Gender Diverse Individuals.\\
    \textbf{Similarity Score:} 71.99\\
    \textbf{Abstract:} This article discusses the impacts of chest and genital gender affirmation surgeries on sexual function for transgender and gender-diverse individuals.
\end{enumerate}

\subsection{Feedback}

After retrieving abstracts, a detailed evaluation was conducted to determine how well the articles aligned with the research question (RQ). The feedback template is meticulously designed to capture and assess the relevance and completeness of the articles. It breaks down the evaluation into specific categories reflecting core components typically assessed in systematic reviews, such as population, intervention, outcomes, and study design. This structured feedback facilitates a thorough and uniform evaluation across different reviewers, ensuring consistency and comprehensiveness. Feedback for the above retrieved articles as provided by domain expert is given in Table \ref{tab:feedback}. 

\begin{table}[!htb]
\centering
\begin{tabular}{p{6cm}p{1.5cm}}
\toprule
\textbf{Feedback Category} & \textbf{Checked} \\
\midrule
Patient / Population / Problem & Yes \\
Intervention / Exposure & Yes \\
Comparison & - \\
Outcome & Yes \\
Study Design / Research Type & Yes \\
Setting / Location & - \\
Phenomenon of Interest & Yes \\
Evaluation & - \\
The article captured all relevant concepts & Yes \\
Other & - \\
\bottomrule
\end{tabular}
\caption{Feedback Form Evaluation of Retrieved Articles}
\label{tab:feedback}
\end{table}

\noindent\textbf{Feedback for the article:} Abstracts 1, 2, 4, and 5 were relevant to the RQ.\\
\noindent\textbf{Computed Relevance Percentage:} 80\%. (for this specific query)

\section{Domain-Specific Resources}\label{appendix:domainmodels}

Retrieval in search systems is performed by scoring the input query and document chunks on a similarity function, usually cosine similarity. The embedding model serves a crucial role in this process. An effective embedding model maps similar input queries and chunks more closely in a semantic space. With the advent of Neural networks, dense retrieval models like the sentence transformers family are now pre-trained on sentence-level general-domain datasets. More recently, embedding models pre-trained on domain-specific data have been developed, particularly for biomedical usage, as they have highly-specialized vocabulary and concepts. 
\cite{kerner2024domain} suggests that domain-specific embedding models perform better than general domain models for tasks requiring specialized domain-specific vocabulary in the corpora. For our experiments, we use both of these kinds of models. Distill-Bert \cite{reimers-2019-sentence-bert}, MPNet \cite{biompnet}, PubMedBert \cite{deka2022improved}, SPECTER \cite{cohan2020specter}, and MedCPT \cite{jin2023medcpt}  are the models used for our experiments. These models' comparative performance is assessed based on relevance percentage, computed based on the evaluation done by domain experts.

Distill-bert and MPNet are sentence-transformer-based embedding models that map input sentences and paragraphs to 768-dimensional dense sentence-level embeddings for downstream use like clustering and semantic search. The MPNet model we use for this experiment is trained on PubMed abstracts and citations. The Distill-bert model specializes in semantic search, having been trained on the MS MARCO dataset. Since our experimental setup involves a single-line input query and paragraph-length paper abstracts, it forms an example of an asymmetric search. The dataset is of a similar format, so the model training is closely related to our experiments. 
SPECTER is a scientific retriever trained on title abstract pairs of scientific publications. PubMedBert takes the BioMedBERT trained on abstracts and full-text articles on PubMed and fine-tunes it over the MS-MARCO Question Answering dataset. MedCPT is a Contrastive Pre-trained Transformer model built for zero-shot biomedical information retrieval. The model is trained on a massive corpus of 255 million query-article pairs from PubMed search logs. The MedCPT Query Encoder computes the embeddings of short texts like search queries, while the MedCPT Article Encoder focuses on encoding biomedical articles. Our overall choice of embedding models covers a multitude of different domains and tasks. 

The comparative evaluation of embedding models for the NeuroLit Navigator system highlighted significant performance differences. To identify the best-performing model, we computed the relevance percentage of retrieved articles with the help of domain experts. MPNet achieved the highest relevance percentage at 36\%, followed by BertDistill at 28\%. SPECTER and PubmedBert also achieve a relevance score of 24\% and 20\%, respectively. Finally, MedCPT becomes the worst performing with a relevance percentage of 4\%. 

\section{Appendix: LLM Outputs} \label{appendix:llmoutputs}

The following sections document outputs from various LLM-based SR tools for specific queries. The top-3 articles from each LLM as of September 13th, 2024, are listed below. Reviewers are asked to evaluate the relevance of each article.

\subsubsection{Query 1: What range of childhood experiences are defined as expanded adverse childhood experiences (ACEs)?}

\paragraph{Scite}
\begin{itemize}
    \item \textbf{Title 1:} Conventional and expanded adverse childhood experiences (ACEs) and maternal functioning among low-income Black mothers [Relevance: \checkbox Yes]
    \item \textbf{Title 2:} Adverse Childhood Experiences and worsened social relationships in adult life among Polish medical and dental students [Relevance: \crossedbox No, as it’s unrelated to expanded ACEs. This article only focuses on conventional ACEs]
    \item \textbf{Title 3:} Development and psychometric properties of the ACE-I: Measuring adverse childhood experiences among Latino immigrant youth. [Relevance: \checkbox Yes]
\end{itemize}

\paragraph{Consensus}
\begin{itemize}
    \item \textbf{Title 1:} Adverse Childhood Experiences: Expanding the Concept of Adversity. [Relevance: \checkbox Yes]
    \item \textbf{Title 2:} Expanding the Original Definition of Adverse Childhood Experiences (ACEs) [Relevance: \checkbox Yes]
    \item \textbf{Title 3:} Unpacking the impact of adverse childhood experiences on adult mental health. [Relevance: \checkbox Yes]
\end{itemize}

\paragraph{Perplexity}
\begin{itemize}
    \item \textbf{Title 1:} About Adverse Childhood Experiences (blog cdc) [Relevance: \crossedbox No, as it is a webpage blog.]
    \item \textbf{Title 2:} Prevalence of Adverse Childhood Experiences Among U.S. Adults — Behavioral Risk Factor Surveillance System, 2011–2020 [Relevance: \crossedbox No, as it misses the expanded concept]
    \item \textbf{Title 3:} The Association Between Expanded ACEs and Behavioral Health Outcomes Among Youth at First Time Legal System Contact [Relevance: \checkbox Yes]
\end{itemize}

\subsubsection{Query 2: Disparities in HIV and STIs diagnoses among children involved in the child welfare system}

\paragraph{Scite}
\begin{itemize}
    \item \textbf{Title 1:} Child Welfare System Involvement Among Children With Medical Complexity [Relevance: \crossedbox No, While it captures the population of interest (children in the child welfare system), it misses either HIV or STI diagnosis]
    \item \textbf{Title 2:} Medicaid Disenrollment Patterns Among Children Coming into Contact with Child Welfare Agencies [Relevance: \crossedbox No, as it captures the population of interest (child in the welfare system) but misses the HIV or STI concept]
    \item \textbf{Title 3:} The Cedar Project: Negative health outcomes associated with involvement in the child welfare system among young Indigenous people who use injection and non-injection drugs in two Canadian cities [Relevance: \checkbox Yes]
\end{itemize}

\paragraph{Consensus}
\begin{itemize}
    \item \textbf{Title 1:} Impact of parent-child communication interventions on sex behaviors and cognitive outcomes for black/African-American and Hispanic/Latino youth: a systematic review, 1988-2012. [Relevance: \crossedbox No, While it captures children and HIV, it misses the child welfare system concept]
    \item \textbf{Title 2:} HIV Among Black Men Who Have Sex with Men (MSM) in the United States: A Review of the Literature [Relevance: \crossedbox No, While it captures the concept of HIV, it’s missing the population (children in the welfare system)]
    \item \textbf{Title 3:} Examining Racial Disparities in HIV: Lessons From Sexually Transmitted Infections Research [Relevance: \crossedbox No, While it captures HIV/STIs, it misses the population (children in the welfare system)]
\end{itemize}

\paragraph{Perplexity}
\begin{itemize}
    \item \textbf{Title 1:} A Report on Infants and Children with HIV Infection in Foster Care [Relevance: \crossedbox No, unless gray literature is of interest.]
    \item \textbf{Title 2:} Foster care, syndemic health disparities and associations with HIV/STI diagnoses among young adult substance users [Relevance: \checkbox Yes]
    \item \textbf{Title 3:} AIDS Children and Child Welfare (blog) [Relevance: \crossedbox No, unless gray literature is of interest.]
\end{itemize}

\bibliography{bigdata}

\begin{thebibliography}{43}
\providecommand{\natexlab}[1]{#1}

\bibitem[{Andersen et~al.(2024)Andersen, Zeinert, Rosenberg, and Fonnes}]{andersen2024comparative}
Andersen, M.~Z.; Zeinert, P.; Rosenberg, J.; and Fonnes, S. 2024.
\newblock Comparative analysis of Cochrane and non-Cochrane reviews over three decades.
\newblock \emph{Systematic Reviews}, 13(1): 120.

\bibitem[{Anonymous(2024)}]{anonymousCoder}
Anonymous, A. 2024.
\newblock Github Repository.
\newblock \url{https://anonymous.4open.science/r/NeuroLitNavigator-6FEE}.
\newblock Accessed: 2024-09-16.

\bibitem[{Aromataris et~al.(2024)Aromataris, Lockwood, Porritt, Pilla, and Jordan}]{JBI2024}
Aromataris, E.; Lockwood, C.; Porritt, K.; Pilla, B.; and Jordan, Z., eds. 2024.
\newblock \emph{JBI Manual for Evidence Synthesis}.
\newblock JBI.

\bibitem[{Bodenreider(2004)}]{bodenreider2004unified}
Bodenreider, O. 2004.
\newblock The unified medical language system (UMLS): integrating biomedical terminology.
\newblock \emph{Nucleic acids research}, 32(suppl\_1): D267--D270.

\bibitem[{Bozada~Jr et~al.(2021)Bozada~Jr, Borden, Workman, Del~Cid, Malinowski, and Luechtefeld}]{bozada2021sysrev}
Bozada~Jr, T.; Borden, J.; Workman, J.; Del~Cid, M.; Malinowski, J.; and Luechtefeld, T. 2021.
\newblock Sysrev: a FAIR platform for data curation and systematic evidence review.
\newblock \emph{Frontiers in Artificial Intelligence}, 4: 685298.

\bibitem[{Bramer et~al.(2018)Bramer, De~Jonge, Rethlefsen, Mast, and Kleijnen}]{bramer2018systematic}
Bramer, W.~M.; De~Jonge, G.~B.; Rethlefsen, M.~L.; Mast, F.; and Kleijnen, J. 2018.
\newblock A systematic approach to searching: an efficient and complete method to develop literature searches.
\newblock \emph{Journal of the Medical Library Association: JMLA}, 106(4): 531.

\bibitem[{Bullers et~al.(2018)Bullers, Howard, Hanson, Kearns, Orriola, Polo, and Sakmar}]{bullers2018takes}
Bullers, K.; Howard, A.~M.; Hanson, A.; Kearns, W.~D.; Orriola, J.~J.; Polo, R.~L.; and Sakmar, K.~A. 2018.
\newblock It takes longer than you think: librarian time spent on systematic review tasks.
\newblock \emph{Journal of the Medical Library Association: JMLA}, 106(2): 198.

\bibitem[{Chandler et~al.(2019)Chandler, Cumpston, Li, Page, and Welch}]{chandler2019cochrane}
Chandler, J.; Cumpston, M.; Li, T.; Page, M.~J.; and Welch, V. 2019.
\newblock Cochrane handbook for systematic reviews of interventions.
\newblock \emph{Hoboken: Wiley}.

\bibitem[{Cheng and Augustin(2021)}]{cheng2021keep}
Cheng, S.; and Augustin, C. 2021.
\newblock Keep a human in the machine and other lessons learned from deploying and maintaining colandr.
\newblock \emph{CHANCE}, 34(3): 56--60.

\bibitem[{Cohan et~al.(2020)Cohan, Feldman, Beltagy, Downey, and Weld}]{cohan2020specter}
Cohan, A.; Feldman, S.; Beltagy, I.; Downey, D.; and Weld, D.~S. 2020.
\newblock Specter: Document-level representation learning using citation-informed transformers.
\newblock \emph{arXiv preprint arXiv:2004.07180}.

\bibitem[{consensus(2024)}]{consensus}
consensus. 2024.
\newblock Consensus AI.
\newblock \url{https://consensus.app/}.
\newblock Accessed: 2024-09-16.

\bibitem[{Deka, Jurek-Loughrey, and Deepak(2022)}]{deka2022improved}
Deka, P.; Jurek-Loughrey, A.; and Deepak, P. 2022.
\newblock Improved Methods To Aid Unsupervised Evidence-Based Fact Checking For Online Health News.
\newblock \emph{Journal of Data Intelligence}, 3(4): 474--504.

\bibitem[{Domenichini et~al.(2024)Domenichini, Chiarello, Giordano, and Fantoni}]{domenichini2024llms}
Domenichini, D.; Chiarello, F.; Giordano, V.; and Fantoni, G. 2024.
\newblock LLMs for Knowledge Modeling: NLP Approach to Constructing User Knowledge Models for Personalized Education.
\newblock In \emph{Adjunct Proceedings of the 32nd ACM Conference on User Modeling, Adaptation and Personalization}, 576--583.

\bibitem[{EPPI(2019)}]{eppi_reviewer_web_beta}
EPPI. 2019.
\newblock {EPPI-Reviewer Web (Beta)}.
\newblock \url{https://eppi.ioe.ac.uk/eppireviewer-web/home}.
\newblock Accessed: 2024-09-16.

\bibitem[{GBaker(2024)}]{biompnet}
GBaker. 2024.
\newblock BioLinkBert MPNet.
\newblock \url{https://huggingface.co/GBaker/biolinkbert-base-medqa-usmle-MPNet-context}.
\newblock Accessed: 2024-09-16.

\bibitem[{Gross, Taylor, and Joudrey(2015)}]{gross2015still}
Gross, T.; Taylor, A.~G.; and Joudrey, D.~N. 2015.
\newblock Still a lot to lose: the role of controlled vocabulary in keyword searching.
\newblock \emph{Cataloging \& classification quarterly}, 53(1): 1--39.

\bibitem[{Howard et~al.(2020)Howard, Phillips, Tandon, Maharana, Elmore, Mav, Sedykh, Thayer, Merrick, Walker et~al.}]{howard2020swift}
Howard, B.~E.; Phillips, J.; Tandon, A.; Maharana, A.; Elmore, R.; Mav, D.; Sedykh, A.; Thayer, K.; Merrick, B.~A.; Walker, V.; et~al. 2020.
\newblock SWIFT-Active Screener: Accelerated document screening through active learning and integrated recall estimation.
\newblock \emph{Environment International}, 138: 105623.

\bibitem[{Jin et~al.(2023)Jin, Kim, Chen, Comeau, Yeganova, Wilbur, and Lu}]{jin2023medcpt}
Jin, Q.; Kim, W.; Chen, Q.; Comeau, D.~C.; Yeganova, L.; Wilbur, W.~J.; and Lu, Z. 2023.
\newblock MedCPT: Contrastive Pre-trained Transformers with large-scale PubMed search logs for zero-shot biomedical information retrieval.
\newblock \emph{Bioinformatics}, 39(11): btad651.

\bibitem[{Kerner(2024)}]{kerner2024domain}
Kerner, T. 2024.
\newblock Domain-Specific Pretraining of Language Models: A Comparative Study in the Medical Field.
\newblock \emph{arXiv preprint arXiv:2407.14076}.

\bibitem[{Koffel(2015)}]{koffel2015use}
Koffel, J.~B. 2015.
\newblock Use of recommended search strategies in systematic reviews and the impact of librarian involvement: a cross-sectional survey of recent authors.
\newblock \emph{PloS one}, 10(5): e0125931.

\bibitem[{Lewis et~al.(2020)Lewis, Perez, Piktus, Petroni, Karpukhin, Goyal, K{\"u}ttler, Lewis, Yih, Rockt{\"a}schel et~al.}]{lewis2020retrieval}
Lewis, P.; Perez, E.; Piktus, A.; Petroni, F.; Karpukhin, V.; Goyal, N.; K{\"u}ttler, H.; Lewis, M.; Yih, W.-t.; Rockt{\"a}schel, T.; et~al. 2020.
\newblock Retrieval-augmented generation for knowledge-intensive nlp tasks.
\newblock \emph{Advances in Neural Information Processing Systems}, 33: 9459--9474.

\bibitem[{Lindberg(1990)}]{lindberg1990unified}
Lindberg, C. 1990.
\newblock The Unified Medical Language System (UMLS) of the National Library of Medicine.
\newblock \emph{Journal (American Medical Record Association)}, 61(5): 40--42.

\bibitem[{Neumann et~al.(2019)Neumann, King, Beltagy, and Ammar}]{neumann-etal-2019-scispacy}
Neumann, M.; King, D.; Beltagy, I.; and Ammar, W. 2019.
\newblock {S}cispa{C}y: {F}ast and {R}obust {M}odels for {B}iomedical {N}atural {L}anguage {P}rocessing.
\newblock In \emph{Proceedings of the 18th BioNLP Workshop and Shared Task}, 319--327. Florence, Italy: Association for Computational Linguistics.

\bibitem[{NIH-NLM(2024)}]{mesh}
NIH-NLM. 2024.
\newblock Welcome to Medical Subject Headings.
\newblock \url{https://www.nlm.nih.gov/mesh/meshhome.html}.
\newblock Accessed: 2024-09-16.

\bibitem[{perplexity(2024)}]{perplex}
perplexity. 2024.
\newblock Perplexity AI.
\newblock \url{https://www.perplexity.ai/}.
\newblock Accessed: 2024-09-16.

\bibitem[{Rawte et~al.(2023)Rawte, Chakraborty, Pathak, Sarkar, Tonmoy, Chadha, Sheth, and Das}]{rawte2023troubling}
Rawte, V.; Chakraborty, S.; Pathak, A.; Sarkar, A.; Tonmoy, S. T.~I.; Chadha, A.; Sheth, A.; and Das, A. 2023.
\newblock The Troubling Emergence of Hallucination in Large Language Models-An Extensive Definition, Quantification, and Prescriptive Remediations.
\newblock In \emph{Proceedings of the 2023 Conference on Empirical Methods in Natural Language Processing}, 2541--2573.

\bibitem[{Reimers and Gurevych(2019)}]{reimers-2019-sentence-bert}
Reimers, N.; and Gurevych, I. 2019.
\newblock Sentence-BERT: Sentence Embeddings using Siamese BERT-Networks.
\newblock In \emph{Proceedings of the 2019 Conference on Empirical Methods in Natural Language Processing}. Association for Computational Linguistics.

\bibitem[{Rethlefsen et~al.(2015)Rethlefsen, Farrell, Trzasko, and Brigham}]{rethlefsen2015librarian}
Rethlefsen, M.~L.; Farrell, A.~M.; Trzasko, L. C.~O.; and Brigham, T.~J. 2015.
\newblock Librarian co-authors correlated with higher quality reported search strategies in general internal medicine systematic reviews.
\newblock \emph{Journal of clinical epidemiology}, 68(6): 617--626.

\bibitem[{Roy et~al.(2024)Roy, Khandelwal, Vera, Surana, Heckman, and Sheth}]{roy2024gear}
Roy, K.; Khandelwal, V.; Vera, V.; Surana, H.; Heckman, H.; and Sheth, A. 2024.
\newblock GEAR-Up: Generative AI and External Knowledge-Based Retrieval: Upgrading Scholarly Article Searches for Systematic Reviews.
\newblock In \emph{Proceedings of the AAAI Conference on Artificial Intelligence}, volume~38, 23823--23825.

\bibitem[{scite(2024)}]{scite}
scite. 2024.
\newblock Scite\_ AI.
\newblock \url{https://scite.ai}.
\newblock Accessed: 2024-09-16.

\bibitem[{Sheth et~al.(2021)Sheth, Gaur, Roy, and Faldu}]{sheth2021knowledge}
Sheth, A.; Gaur, M.; Roy, K.; and Faldu, K. 2021.
\newblock Knowledge-intensive language understanding for explainable ai.
\newblock \emph{IEEE IC}, 25(5): 19--24.

\bibitem[{Sheth et~al.(2022)Sheth, Gaur, Roy, Venkataraman, and Khandelwal}]{sheth2022process}
Sheth, A.; Gaur, M.; Roy, K.; Venkataraman, R.; and Khandelwal, V. 2022.
\newblock Process knowledge-infused ai: Toward user-level explainability, interpretability, and safety.
\newblock \emph{IEEE Internet Computing}, 26(5): 76--84.

\bibitem[{Sheth, Pallagani, and Roy(2024)}]{sheth2024neurosymbolicb}
Sheth, A.; Pallagani, V.; and Roy, K. 2024.
\newblock Neurosymbolic ai for enhancing instructability in generative ai.
\newblock \emph{IEEE Intelligent Systems}, 39(5): 5--11.

\bibitem[{Sheth and Roy(2024)}]{sheth2024neurosymbolic}
Sheth, A.; and Roy, K. 2024.
\newblock Neurosymbolic value-inspired artificial intelligence (why, what, and how).
\newblock \emph{IEEE Intelligent Systems}, 39(1): 5--11.

\bibitem[{Sheth, Roy, and Gaur(2023)}]{sheth2023neurosymbolic}
Sheth, A.; Roy, K.; and Gaur, M. 2023.
\newblock Neurosymbolic artificial intelligence (why, what, and how).
\newblock \emph{IEEE IS}, 38(3): 56--62.

\bibitem[{Sheth et~al.(2024)Sheth, Roy, Purohit, and Das}]{sheth2024civilizing}
Sheth, A.; Roy, K.; Purohit, H.; and Das, A. 2024.
\newblock Civilizing and humanizing artificial intelligence in the age of large language models.
\newblock \emph{IEEE Internet Computing}, 28(5): 5--10.

\bibitem[{Sheth et~al.(2025)Sheth, Roy, Venkataramanan, and Nadimuthu}]{sheth2025c}
Sheth, A.~P.; Roy, K.; Venkataramanan, R.; and Nadimuthu, V. 2025.
\newblock C 3 AN: Custom, Compact and Composite AI Systems-A NeuroSymbolic Approach: 4 th-Generation Evolution of Intelligent Systems.

\bibitem[{Thomas et~al.(2021)Thomas, McDonald, Noel-Storr, Shemilt, Elliott, Mavergames, and Marshall}]{thomas2021machine}
Thomas, J.; McDonald, S.; Noel-Storr, A.; Shemilt, I.; Elliott, J.; Mavergames, C.; and Marshall, I.~J. 2021.
\newblock Machine learning reduced workload with minimal risk of missing studies: development and evaluation of a randomized controlled trial classifier for Cochrane Reviews.
\newblock \emph{Journal of Clinical Epidemiology}, 133: 140--151.

\bibitem[{Tilwani et~al.(2024)Tilwani, Saxena, Mohammadi, Raff, Sheth, Parthasarathy, and Gaur}]{tilwani2024reasons}
Tilwani, D.; Saxena, Y.; Mohammadi, A.; Raff, E.; Sheth, A.; Parthasarathy, S.; and Gaur, M. 2024.
\newblock REASONS: A benchmark for REtrieval and Automated citationS Of scieNtific Sentences using Public and Proprietary LLMs.
\newblock \emph{arXiv preprint arXiv:2405.02228}.

\bibitem[{Walker et~al.(2022)Walker, Schmitt, Wolfe, Nowak, Kulesza, Williams, Shin, Cohen, Burch, Stout et~al.}]{walker2022evaluation}
Walker, V.~R.; Schmitt, C.~P.; Wolfe, M.~S.; Nowak, A.~J.; Kulesza, K.; Williams, A.~R.; Shin, R.; Cohen, J.; Burch, D.; Stout, M.~D.; et~al. 2022.
\newblock Evaluation of a semi-automated data extraction tool for public health literature-based reviews: Dextr.
\newblock \emph{Environment international}, 159: 107025.

\bibitem[{Wang et~al.(2023)Wang, Liu, Ying, Yang, Chen, Liu, Zhang, Yan, Lu, Gao et~al.}]{wang2023optimized}
Wang, G.; Liu, X.; Ying, Z.; Yang, G.; Chen, Z.; Liu, Z.; Zhang, M.; Yan, H.; Lu, Y.; Gao, Y.; et~al. 2023.
\newblock Optimized glycemic control of type 2 diabetes with reinforcement learning: a proof-of-concept trial.
\newblock \emph{Nature Medicine}, 29(10): 2633--2642.

\bibitem[{Yang et~al.(2024)Yang, Rao, Chen, Guo, Zhang, Yang, and Zhang}]{yang2024rag}
Yang, D.; Rao, J.; Chen, K.; Guo, X.; Zhang, Y.; Yang, J.; and Zhang, Y. 2024.
\newblock IM-RAG: Multi-Round Retrieval-Augmented Generation Through Learning Inner Monologues.
\newblock In \emph{Proceedings of the 47th International ACM SIGIR Conference on Research and Development in Information Retrieval}, 730--740.

\bibitem[{Yu and Menzies(2019)}]{yu2019fast2}
Yu, Z.; and Menzies, T. 2019.
\newblock FAST2: An intelligent assistant for finding relevant papers.
\newblock \emph{Expert Systems with Applications}, 120: 57--71.

\end{thebibliography}

\end{document}